\documentclass[aps,prl,showpacs,twocolumn,amsmath,amssymb]{revtex4}
\usepackage{graphicx}
\usepackage{epsfig,color}
\usepackage{bm}
\usepackage{amssymb}
\usepackage{amsmath}
\usepackage{latexsym}
\usepackage{longtable}

\newcommand{\vc}[1]{\mbox{\boldmath $#1$}} 
\newcommand{\ind}[1]{_{#1}}    
\newcommand{\indrm}[1]{_{\mathrm {#1}}}    
\newcommand{\dhkl}{d_{\ind{H}}}    

\newcommand{\dei}[1]{\Delta E_{\ind{#1}}^{({\mathrm s})}}   
\newcommand{\dai}[1]{\Delta \theta_{\ind{#1}}^{({\mathrm s})}}   
\newcommand{\cddw}{CD\filter DW}   
\newcommand{\mono}{M}   
\newcommand{\ana}{A}   
\newcommand{\filter}{F} 
\newcommand{\cdaadw}{CDF$_{\mathrm  r}$F$_{\mathrm b}$DW}   
\newcommand{\cdaa}{CDF$_{\mathrm  r}$F$_{\mathrm b}$}   
\newcommand{\cw}{CFW}   
\newcommand{\adat}{AD\&AT}   
\newcommand{\rotangleone}{\tilde{\Theta}_{\indrm{D_1}}}   
\newcommand{\rotangletwo}{\tilde{\Theta}_{\indrm{D_2}}}   

\begin{document}
%
\title{Angular Dispersion and Anomalous Transmission Cast Ultramonochromatic X~Rays} 
\author{Yuri Shvyd'ko} \affiliation{Advanced Photon Source,
  Argonne National Laboratory, Argonne, Illinois 60439, USA}
\author{Stanislav Stoupin} \affiliation{Advanced Photon Source,
  Argonne National Laboratory, Argonne, Illinois 60439, USA}
\author{Deming Shu} \affiliation{Advanced Photon Source,
  Argonne National Laboratory, Argonne, Illinois 60439, USA}
\author{Ruben Khachatryan} \affiliation{Advanced Photon Source,
  Argonne National Laboratory, Argonne, Illinois 60439, USA}
\date{\today}

\begin{abstract}
  Optical spectrometers, instruments that work with monochromatic
  light, are commonly rated by the spectral bandwidth, which defines
  the ability to resolve closely spaced spectral components. The
  ability to detect faint objects among these components, spectral
  contrast, is another desired aspect.  Here we demonstrate that a
  combined effect of angular dispersion (AD) and anomalous
  transmission (AT) of x~rays in Bragg reflection from asymmetrically
  cut crystals can shape spectral distributions of x~rays to profiles
  with record high contrast and small bandwidths.

  The \adat\ x-ray optics is implemented as a five-reflection
  three-crystal arrangement featuring a combination of the above
  mentioned attributes, so much desirable for x-ray monochromators and
  analyzers: a spectral contrast of $\simeq 500$, a bandwidth of
  $\simeq 0.46$~meV and a remarkably large angular acceptance of
  $\simeq 107~\mu$rad.  The new optics can become a foundation for the
  next generation inelastic x-ray scattering spectrometers for studies
  of atomic dynamics.

\end{abstract}




\maketitle
\section{Notations}
Notations and their definitions are listed below in order of
appearance. They are similar to those used in \cite{Shvydko-SB}.
\setlongtables
\begin{longtable}{lp{7cm}}
  $E$ & Photon energy.\\
  $E_{\ind{0}}$ & Average or peak photon energy.\\
  $\Delta E$ & Spectral bandwidth.\\
  $\dei{H}$ & Intrinsic spectral width in symmetric Bragg diffraction.\\
  $\Delta E_{\indrm{\mono}}$ & Bandwidth of the monochromator spectral resolution function (FWHM).\\
  $\eta, \eta_{\ind{H}}$ & Asymmetry angle - angle between reflecting atomic planes and crystal face.\\
  $\theta, \theta_{\ind{H}}$ & Glancing angle of incidence.\\
  $\theta^{\prime}$, $\theta^{\prime}_{\ind{H}}$ & Glancing angle of reflection.\\
  $\Psi$ & Offset between the angle of anomalous transmission and Bragg reflection peaks.\\
  $\Theta$ & Angle of incidence $\Theta=\pi/2-\theta$.\\
  $E_{\indrm{R}}$ & Center energy of the region of exact Bragg back-reflection.\\
  $\Theta_{\indrm{R}}$ & Center incidence angle of the region of exact Bragg back-reflection.\\
  $\dhkl$ & spacing between the reflecting atomic planes.\\
  $h$ & Planck's constant.\\
  $c$ & Speed of light in vacuum.           \\
  $f_{\indrm{\mono}}(E)$ & Monochromator spectral resolution  function.\\
  $C_{\indrm{\mono}}$ &  Contrast of the spectral resolution function.\\
  $\Delta  E_{\indrm{C}}$ & Energy offset at which spectral contrast is measured.\\
  $\vc{H}$ & Diffraction  vectors: $\vc{H}=\vc{C}$,$\vc{D}_1$,$\vc{\filter }$,$\vc{D}_2$, or $\vc{W}$.\\
  C  & Collimator.\\
  D & Dispersing element.\\
  \filter & Anomalous transmission filter.\\
  W & Wavelength selector.\\
  $(hkl)$ & Miller indices of the diffraction vector $\vc{H}$.\\
  $b_{\ind{H}}$ & Asymmetry parameter.\\
  $\dai{H} $ & Intrinsic angular width in symmetric diffraction.\\
  $w_{\ind{H}}^{({\mathrm s})}$ & Refraction correction in symmetric  diffraction.\\
  $d$ & Crystal thickness.\\
  $\lambda$ & X-ray wavelength.\\
  $\lambda_{\indrm{R}}$& Center wavelength of the region of exact Bragg back-reflection.\\
  $\Delta \theta_{\indrm{\mono  }}$ & Angular acceptance  of the monochromator.\\
  $\Delta\theta_{\indrm{\mono }}^{\prime}$ & Angular divergence of x~rays emanating from the monochromator.\\
  $\Delta\theta_{\indrm{H}}$ & Angular width of a generic Bragg reflection.\\
  $\Delta\theta_{\ind{H}}^{\prime}$ &  Angular width of reflected x~rays for a generic  Bragg reflection.\\
  $\tilde\Theta_{\indrm{D}}$ & Rotation angle of D-crystals.\\
  $\Delta E_{\indrm{\mono \otimes \ana}}$ & Bandwidth of the combined
  spectral resolution functions of the
  monochromator and  analyzer.\\
  $\Delta E_{_{\indrm{1/10000}}}$ & Spectral half width at the $10^{-4}$
  level fraction
  of the maximum.\\
  $\varepsilon_{\indrm{\mono }}$ & Average  spectral efficiency of the monochromator.\\
  $I_{\indrm{0}}$ &  Incident photon flux.\\
  $I_{\indrm{\mono }}$ & Photon flux transmitted through the monochromator.\\
  $\Delta  E_{\indrm{0}}$ & Bandwidth of the  spectral distribution of the incident radiation.\\
  $\theta_{\indrm{M}}$ & Rotation angle of the monochromator.\\
  $\theta_{\indrm{A}}$ & Rotation angle of the analyzer.
\end{longtable}

\section{Introduction}

Despite many recent advances in inelastic x-ray and neutron scattering
critical voids exist in current experimental capabilities for
investigation of atomic dynamics in biomaterials (DNA, lipid bilayers,
proteins), in many intriguing classes of oxide materials (high
temperature superconductors, colossal magnetoresistance manganites,
multiferroics), and many other materials with diverse properties of
fundamental and practical interest. This void calls for new hard x-ray
spectrometers capable of not only achieving small spectral bandwidths
$\Delta E$ in the 0.1-1 meV range ($\Delta E/E\approx
10^{-7}-10^{-8}$), but, more importantly, the ability to detect faint
spectral objects, which requires small bandwidth at the
$10^{-3}-10^{-4}$ level fraction of the spectral resolution function
maximum.  In this paper, we present a new concept for achieving highly
monochromatic x~rays with steeply declining tails (large spectral
contrast) as well as its realization.

\begin{figure}[t!]
\setlength{\unitlength}{\textwidth}
\begin{picture}(1,0.21)(0,0)
\put(0.0,0.00){\includegraphics[width=0.50\textwidth]{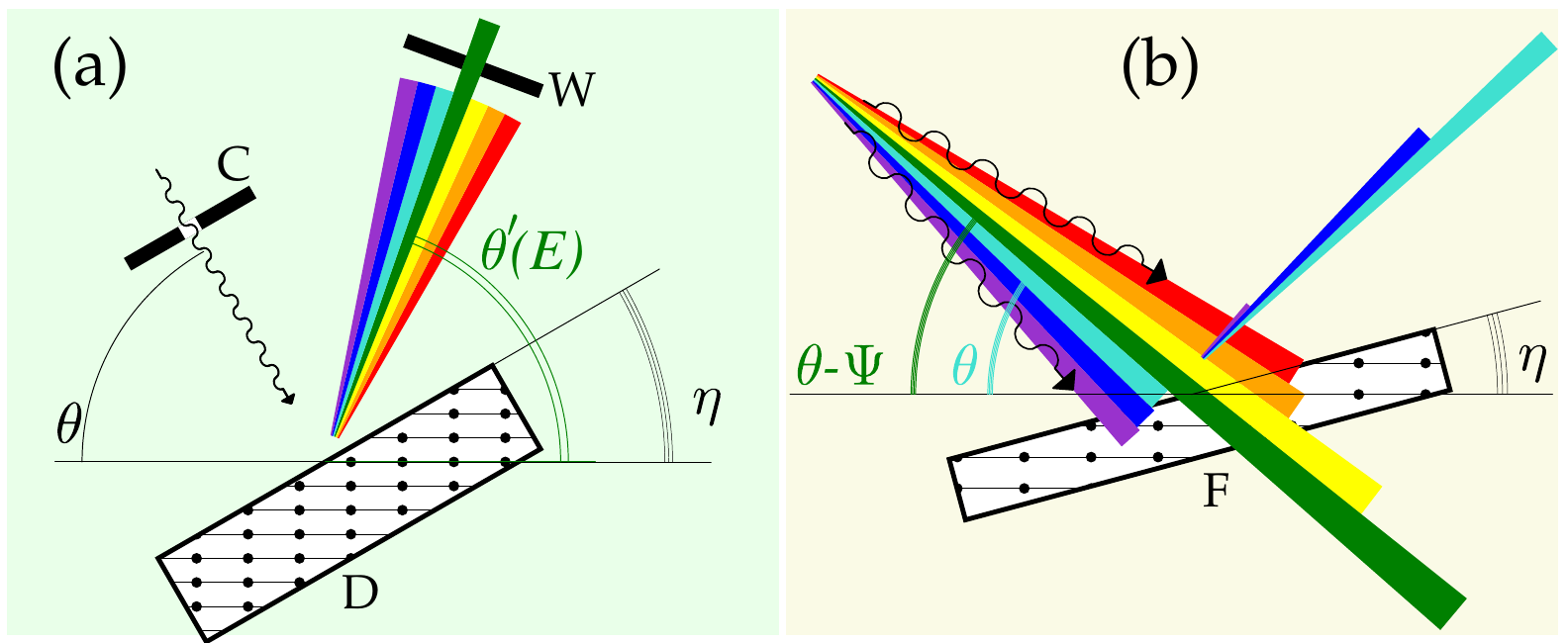}}
\end{picture}
\caption{Basic phenomena underlying the \adat\ x-ray optics. In x-ray
  Bragg diffraction from atomic planes composing nonzero angle $\eta$
  to the crystal entrance face, the crystal acts (a) like an optical
  prism dispersing the photons into a divergent x-ray fan with photons
  of different energies $E$ propagating at different reflection angles
  $\theta^{\prime}(E)$ - effect of angular dispersion (AD)
  \cite{KB76,BSS95,Shvydko-SB,SLK06}, (b) as a filter with anomalously
  high transparency for x~rays with incidence angles $\theta-\Psi$ in
  the immediate vicinity ($\Psi \approx 5~\mu$rad) but smaller than
  the Bragg angle $\theta$ - effect of anomalous transmission (AT)
  \cite{Wagner56,Kishino71,BHS75}.}
\label{fig001} 
\end{figure}

Principles of monochromatization of hard x~rays in essence are based
on Bragg diffraction of x~rays from periodic gratings of atomic planes
in single crystals (for review and references see, e.g.,
\cite{Shvydko-SB}). Spectral band in which x~rays are reflected, the
Bragg diffraction intrinsic width $\dei{H}$, is typically small, not
more than $\dei{H} /E_{\ind{0}} \simeq 10^{-4}$ if measured relative
to an average photon energy $E_{\ind{0}}$.  The smallness of $\dei{H}
$ is determined first of all by a macroscopically large number of
reflecting atomic planes, as well as by crystal and atomic properties.
The intrinsic Bragg bandwidth can be reduced
\cite{ChM96,THS97,Toellner00,CRL01,Yabashi01,Toellner01,TAS06,TAS11}
by using the so-called asymmetric x-ray diffraction, diffraction from
atomic planes at nonzero angle $\eta$ to the crystal
face - Fig.~\ref{fig001}. Still, the bandwidth cannot be tailored to
arbitrary small values without significant loss in Bragg reflectivity.

The intrinsic Bragg reflection width $\dei{H}$ basically sets the
limit for the smallest band in which x~rays can be selected with a
given Bragg reflection, and, therefore, sets the limit for
monochromatization of x~rays.  This fundamental limitation, can be
overcome if an effect of angular dispersion in Bragg diffraction from
asymmetrically cut crystals \cite{KB76,BSS95,Shvydko-SB} is used, as
proposed in \cite{Shvydko-SB}. Bragg diffraction of x~rays from
asymmetrically cut crystals have the same effect on x~rays as an
optical prism on visible light: an incident collimated x-ray beam is
fanned-out upon reflection with photons of different energies
propagating at different angles $\theta^{\prime}(E)$ -
Fig.~\ref{fig001}(a), with a dispersion rate
\begin{equation}
  \frac{{\mathrm d} \theta^{\prime}}{{\mathrm d} E}\, =\, \frac{2}{E}\,\frac{\sin\theta \sin\eta}{\sin(\theta-\eta)} \,
\xrightarrow{\,\,\theta \rightarrow 90^{\circ}\,\,}\, 
\frac{2\tan\eta}{E},
\label{eq001}
\end{equation}
as demonstrated in \cite{SLK06}.  By picking out photons in a small
angular range $\Delta\theta^{\prime}$ from the fan, the bandwidth
$\Delta E_{\indrm{\mono}}$ of the selected x~rays can be reduced to
any small value, independent of how large is the intrinsic Bragg
reflection width $\dei{H}$.  In the proof of the principle experiments
\cite{SLK06,SKR06} it was confirmed that the angular dispersion indeed
can be used to overcome the Bragg reflection width limitation, though,
no spectacular small bandwidths has been achieved until now. In this
paper we introduce an advanced x-ray optics with enhanced angular
dispersion, allowing to monochromatize x~rays to bandwidths $\Delta
E_{\indrm{\mono}}/\dei{H}\simeq 10^{-2}$, which is almost two orders
of magnitude smaller than the relative intrinsic bandwidth of the
applied Bragg reflection.

However, achieving steep tails of the spectral function, i.e., high
spectral contrast\footnote{The contrast of the spectral resolution
  function $f_{\indrm{\mono}}(E)$, is defined as a ratio
  $C_{\indrm{\mono}}=f(0)/f(\Delta E_{\indrm{C}})$ of the peak
  intensity to the intensity of the tail at an offset $E=\Delta
  E_{\indrm{C}}=1.578\,\Delta E_{\indrm{\mono}}$. Here $\Delta
  E_{\indrm{\mono}}$ is the bandwidth of $f_{\indrm{\mono}}(E)$.  The
  spectral contrast is thus a dynamic range available to resolve a
  faint feature with an energy resolution equal to $\Delta
  E_{\indrm{C}}$.  The offset $\Delta E_{\indrm{C}}$ is chosen such
  that the contrast of the Gaussian function
  $C_{\indrm{\mono}}=10^3$. For the Lorentzian function it is two
  orders of magnitude smaller: $C_{\indrm{\mono}}=11$.}, is the most
challenging task in monochromatization of x rays.
Steep tails are equally important in spectroscopic
applications.
Dynamical theory of x-ray Bragg diffraction in crystals predicts that
the tails of the spectral reflection function decline slowly $\propto
1/(E-E_0)^2$, for a single Bragg reflection.  While suppression of the
tails can be achieved using a sequence of Bragg reflections, a more
dramatic improvement is demonstrated here.  Tails as steep as those of
the Gaussian distribution can be obtained with a qualitatively new
approach based on the effect of anomalous transmission of x~rays in
Bragg diffraction from asymmetrically cut crystals
\cite{Wagner56,Kishino71,BHS75}.  In fact, it is a combination of
angular dispersion and anomalous transmission which yields the
extremely steep tails and a narrow bandwidth.  The effect of angular
dispersion due to an asymmetric Bragg reflection is illustrated in
Fig.~\ref{fig001}(a).  The combined effect is illustrated in
Fig.~\ref{fig001}(b).  A part of the dispersion fan with glancing
angles of incidence $\theta-\Psi$ in the immediate vicinity ($\Psi
\approx 5~\mu$rad) but smaller than the Bragg angle $\theta$
propagates through the crystal with anomalously low absorption, while
the rest of the dispersion fan is abruptly rejected by the Bragg
reflection.

A novel x-ray optics introduced here is based on the phenomena of
angular dispersion and anomalous transmission. Particularly, we
describe the underlying principles, design and performance of a
three-crystal five-reflection monochromator, featuring a combination
of superlative properties, such as, exceptionally steep tails of the
spectral profile, an extremely narrow bandpass, an extraordinary large
angular acceptance, high efficiency, and the in-line configuration
(i.e., the incident and the resulting monochromatic x rays are
parallel and propagate in the same direction). The small bandpass is
due to angular dispersion \cite{KB76,BSS95,Shvydko-SB,SLK06} and the
steep tails are due to anomalous transmission
\cite{Wagner56,Kishino71,BHS75}.

\begin{figure}[h!]
\setlength{\unitlength}{\textwidth}
\begin{picture}(1,0.72)(0,0)
\put(0.0,0.00){\includegraphics[width=0.5\textwidth]{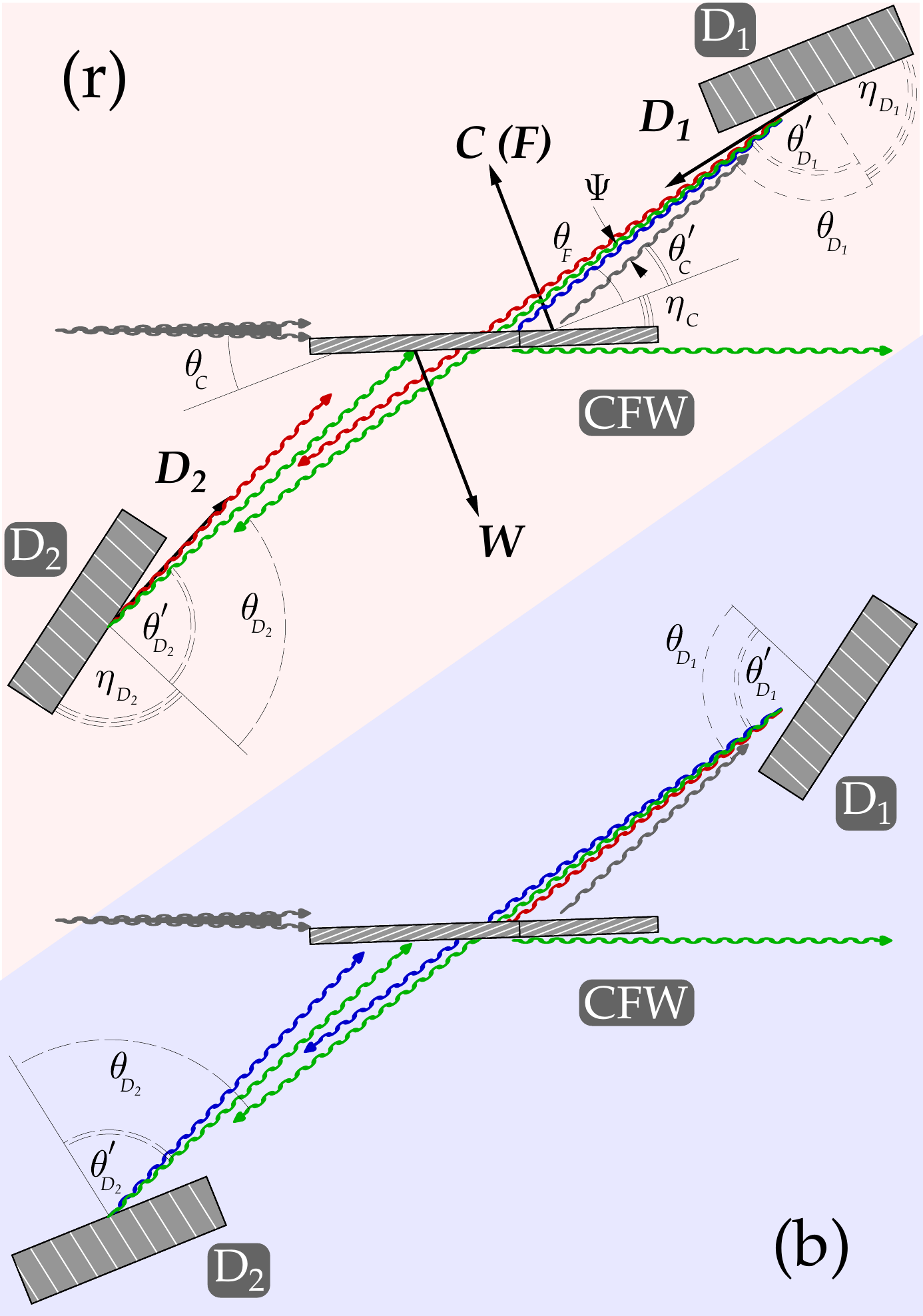}}
\end{picture}
\caption{Optical scheme of the in-line three-crystal five-reflection
  \cddw\ monochromators (r) red-winged and (b) blue-winged,
  respectively. 
  The \cw\ crystal, executes three key functions, a collimator -C, an
  anomalous transmission filter -\filter , and a wavelength selector
  -W in successive reflections.  The crystals D$_1$ and D$_2$ are
  dispersing elements. All crystals are asymmetrically cut with the
  reflecting atomic planes shown by the white lines, perpendicular to
  the diffraction vectors $\vc{H}$ ($\vc{H}=\vc{C}$, $\vc{\filter }$,
  $\vc{D}_1$, $\vc{D}_2$, or $\vc{W}$), composing non-zero asymmetry
  angle $\eta_{\ind{\mathrm H}}$ to the entrance
  surface. $\theta_{\ind{\mathrm H}}$ and $\theta_{\ind{\mathrm
      H}}^{\prime}$ are the glancing angles of incidence and
  reflection, respectively.}
\label{fig002}
\end{figure}

\section{\adat\ Optics Principles  and Implementation}

X-ray monochromators which employ the effect of angular dispersion,
require three fundamental optical elements, each performing a distinct
key function, as schematically shown in Fig.~\ref{fig001}(a).  First, a
collimator (C-element) accepts x~rays with a large angular spread and
collimates into a beam with a small angular divergence.  Secondly, a
dispersing element (D-element) spreads the collimated beam by means of
an asymmetric Bragg reflection into a fan with different spectral
components propagating at different angles. Thirdly, a wavelength
selector (W-element) selects photons from the fan in a small angular,
and, therefore, spectral range \cite{Shvydko-SB}.  An angular
dispersive monochromator with each of the three elements represented
by an individual crystal, a CDW monochromator, have been demonstrated
in \cite{SKR06}. 

Here we introduce novel x-ray optics with a combined effect of angular
dispersion (AD) and anomalous transmission (AT) to produce x~rays with
a spectral distribution having both very steep tails and small
bandwidths.  The \adat\ optics contains an additional element, the
anomalous transmission filter (\filter -element). The \adat\ optics is
realized here by three crystals executing five successive reflections
with key functions C, D, \filter, D, and W respectively (the scheme
was first proposed by Yu.~Shvyd'ko in \cite{NSLS2CDR}). Therefore, in
the following it is termed as \cddw\ monochromator. The four key
functions C,D,W, and \filter , are performed by three crystals: \cw ,
D$_1$, and D$_2$, in two symmetric but nonequivalent configurations:
termed hereafter red-winged - Fig.~\ref{fig002}(r), and blue-winged -
Fig.~\ref{fig002}(b)\footnote{Anomalous transmission was embedded in
  the original proposal of the CDW scheme \cite{Shvydko-SB}, therefore
  in the present terminology it should be defined as CDFW. However the
  \cddw\ scheme is more advantageous mostly due to doubled dispersion
  rate, and in-line geometry}.


The \cw -crystal is a thin asymmetrically cut crystal combining C-,
\filter -, and W-functions.  Incident x~rays with a wide angular
divergence are collimated to a beam with a small divergence upon the
first asymmetric Bragg reflection at glancing angle of reflection
$\theta_{\indrm{C}}^{\prime}$ from the \cw -crystal. The collimated
beam impinges then on the dispersing element D$_1$ at a glancing angle
of incidence $\theta_{\indrm{D_1}}$ in almost exact backscattering
$\theta_{\indrm{D_1}}^{\prime}-\theta_{\indrm{D_1}} =\Psi$. The
asymmetry angle $\eta_{\indrm{D_1}}$ is chosen close to $90^{\circ}$.
The proximity to exact backscattering and $\eta_{\indrm{D_1}} \Rightarrow 90^{\circ}$
are important, first, to ensure the largest effect of angular
dispersion \eqref{eq001} and, secondly, to minimize blurring of the
angular dispersion contrast, which may arise due to the angular spread
of x~rays incident onto the D-crystal \cite{Shvydko-SB}.  The
collimated incident x-ray beam is fanned-out upon reflection from
D$_1$ with photons of different energies propagating towards the \cw\
crystal at different glancing angles of reflection
$\theta_{\indrm{D_1}}^{\prime}(E)$.  The \cw\ crystal now acts as the
\filter -element.  Transmission of photons which impinge upon the
crystal at an angle $\theta_{\indrm{\filter
  }}=\theta^{\prime}_{\indrm{C}}-\Psi$ is anomalously enhanced. This
angle of anomalous transmission is smaller than the Bragg reflection
angle by $\Psi\approx 5~\mu$rad (cf. Fig.~\ref{fig001}).  In the next
step, x~rays are reflected from crystal D$_2$ in the same fashion as
from D$_1$. The resulting angular dispersion rate is that of the
single reflection (Eq.~\eqref{eq001}) increased by a factor of two,
i.e.
\begin{equation}
\frac{{\mathrm d} \theta^{\prime}_{\indrm{D_2}}}{{\mathrm d} E}\,=\,\frac{4\tan\eta_{\indrm{D_2}}}{E}. 
\label{eq004}
\end{equation}
In the final, fifth reflection, the \cw\ crystal in the W-function
selects x~rays in a small angular, and, therefore, spectral range
\footnote{The coplanar scattering geometries shown in
  Fig.~\ref{fig002}, are not mandatory. Wave-vectors of x~rays
  reflected from \cw - and D-crystals may have components
  perpendicular to the scattering plane in the presented
  scheme. Non-coplanar scattering geometry adds a possibility of
  tuning the photon energy in broader range and mitigate the adverse
  effect of multiple reflections in exact backscattering from
  D-crystals \cite{SSSK11}. However, this may compromise the spectral
  resolution, as discussed in \cite{Shvydko-SB}}.

Those photons are preferentially transmitted through the
monochromator, whose energy $E$ and angle of incidence
$\theta_{\indrm{D}}$ to D-crystals are related by the condition of
exact backscattering (a small angular offset $\Psi/2$ is neglected),
because only such photons are also transmitted through the \cw
-crystal . Due to angular dispersion, exact backreflection from an
asymmetrically cut crystal takes place, unlike symmetric diffraction
case, for each photon energy $E$ at different angular deviation
$\Theta=\pi/2-\theta$ from normal incidence to the reflecting atomic
place, as shown in \cite{Shvydko-SB,SLK06}. The relation between the
angle of incidence $\Theta$ and photon energy $E$ for exact
backscattering is given by
\begin{multline}
\Theta-\Theta_{\ind{R}}\, =\, \frac{E-E_{\ind{R}}}{E_{\ind{R}}}\,\tan\eta_{\indrm{H}}, \\
\Theta_{\ind{R}}\, = \,w_{\ind{H}}^{({\mathrm s})} \tan\eta_{\indrm{H}} , 
\hspace{1cm} 
E_{\ind{R}}\,=\,\frac{hc}{2\dhkl} \,(1+w_{\ind{H}}^{({\mathrm s})}).
\label{eq003}
\end{multline}
with $\Theta_{\ind{R}}$ as the angle of incidence and $E_{\ind{R}}$ as
the photon energy in the center of the Bragg reflection region,
$w_{\ind{H}}^{({\mathrm s})}$ is Bragg's reflection refraction
correction, and $\dhkl$ is distance between the reflecting atomic
planes associated with the reciprocal vector $\vc{H}$.  This relation
suggest that by changing simultaneously the angles of incidence to
D-crystals, the energy tuning of the \cddw\ monochromator can be
achieved.  An angular variation of $\delta\Theta_{\indrm{D}}$
according to Eq.~\eqref{eq003} results in a photon energy variation
\begin{equation}
\delta E\,=\,E_{\ind{R}}\,\frac{\delta\Theta_{\indrm{D}}}{\tan\eta_{\indrm{D}}}. 
\label{eq005}
\end{equation}

\begin{figure}[t!]
\setlength{\unitlength}{\textwidth}
\begin{picture}(1,0.80)(0,0)
\put(0.03,0.0){\includegraphics[width=0.43\textwidth]{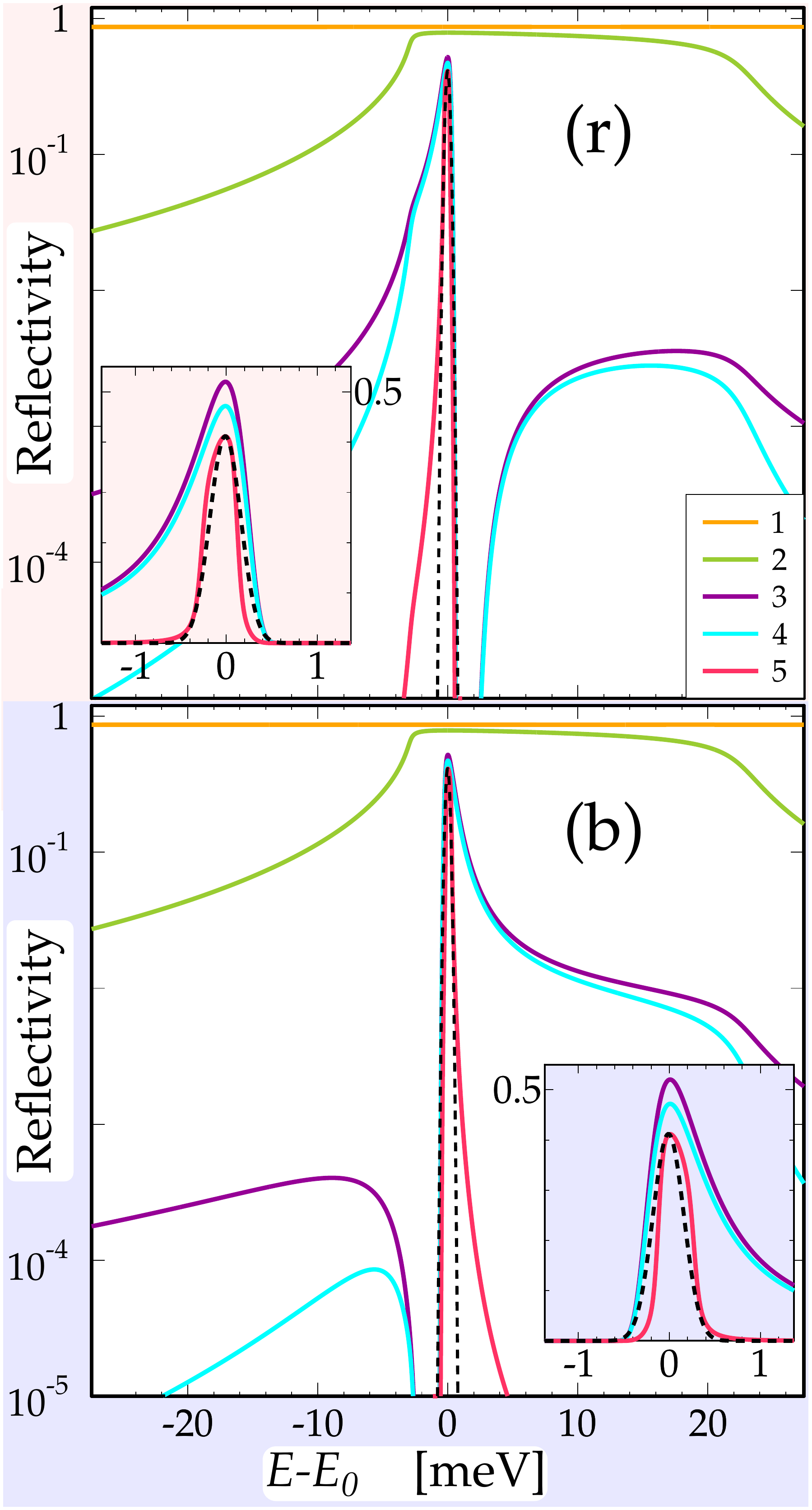}}
\end{picture}
\caption{Dynamical theory calculations of the
spectral distribution of  x~rays after each successive reflection (indicated by number
  and color) from the crystals of the \cddw\ optics in the red-winged
  - Fig.~\ref{fig002}(r), and in the blue-winged configuration -
  Fig.~\ref{fig002}(b), respectively. Black dashed lines show Gaussian distribution of the same 
full width at half maximum. Insets show the distributions on the linear scale. 
}
\label{fig003} 
\end{figure}

\addtocounter{table}{-1} 
\begin{table}[t!]
\centering
\begin{tabular}{|l|llllllll|}
  \hline 
  crystal/ &$\vc{H}$ &$\eta_{\ind{H}} $ &$\theta_{\ind{H}} $ & $b_{\ind{H}}$ & $\dei{H} $ &  $\dai{H}$   & $w_{\ind{H}}^{({\mathrm s})}$ &  $d$\\[-5pt]    
  function &         &     &      &              &     &          &  &       \\
  \cline{2-9}
  & $(hkl)$ & [deg] & [deg]  &  & [meV]  &  [$\mu$rad] & $\times 10^{-6}$ & [mm] \\[0pt]    
  \hline  
  \cw /C        & (2~2~0) &  19.0  &  20.7  & -0.047  & 565 & 23.5 & 47.2  &0.3  \\[-0.0pt]
  D$_1$/D        & (8~0~0) &  88.0  &  89.9  & -1  & 27 &  1870 & 9.13  & 20 \\[-0.0pt]
  \cw /F        & (\={2}~\={2}~0) &  19.0  &  20.7  & -21.5  & 565 &  23.5 & 47.2 & 0.3 \\[-0.0pt]
  D$_2$/D        & (8~0~0) &  88.0  &  89.9  & -1  & 27 &  1870 & 9.13 & 20\\[-0.0pt]
  \cw /W        & (2~2~0) &  19.0  &  20.7  & -21.5  & 565 &  23.5 & 47.2 & 0.3 \\[-0.0pt]
  \hline  
\end{tabular}
\caption{Elements of the \cddw\ optics, and their crystal, and Bragg reflection parameters 
  as used in all presented here dynamical theory calculations and in the experiment:
  $(hkl)$ - Miller indices of the Bragg diffraction vector $\vc{H}$,  
  $\eta_{\ind{H}}$ - asymmetry angle, $\theta_{\ind{H}}$ - glancings angle of incidence, 
  $b_{\ind{H}}=-\sin(\theta_{\ind{H}}\pm\eta_{\ind{H}})/\sin(\theta_{\ind{H}}\mp\eta_{\ind{H}})$ - 
  asymmetry parameter, $d$ - crystal thickness,  $\dai{H}$, $\dei{H}$, and $w_{\ind{H}}^{({\mathrm s})}$  - 
  are Bragg's reflection intrinsic spectral width, angular acceptance, and refraction correction 
  in symmetric scattering geometry, respectively. X-ray photon energy $E=9.1315$~keV.
}
\label{tab1}
\end{table}

\begin{figure*}[t!]
\setlength{\unitlength}{\textwidth}
\begin{picture}(1,0.28)(0,0)
\put(0.10,0.00){\includegraphics[width=0.8\textwidth]{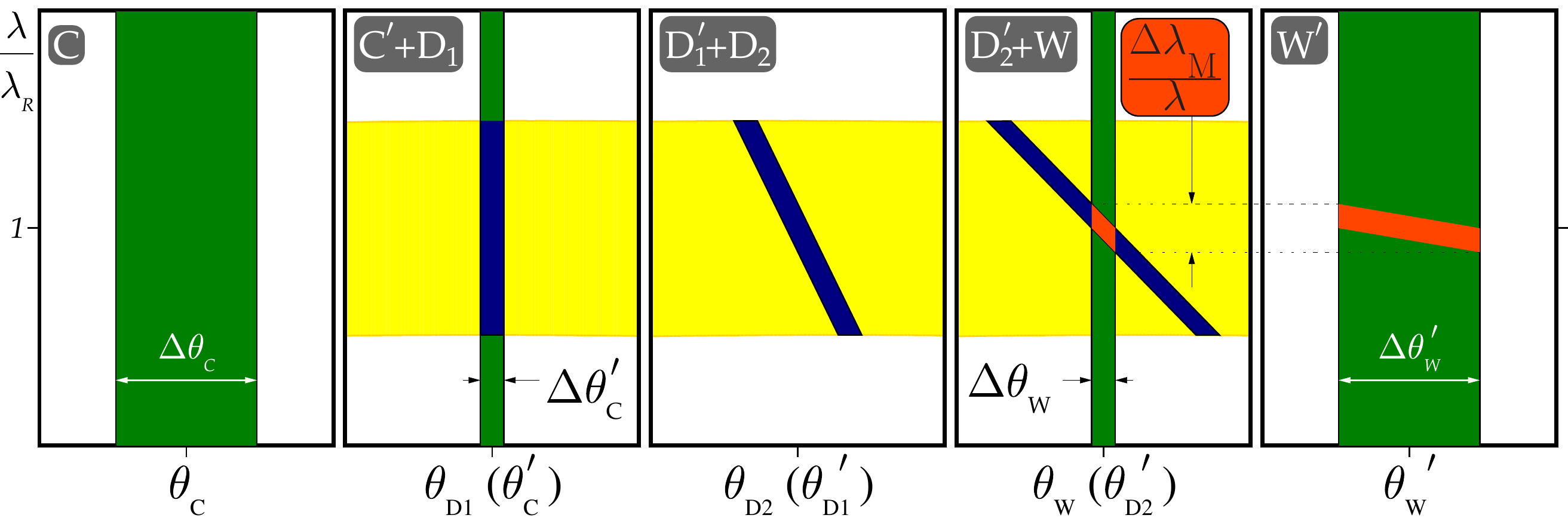}}
\end{picture}
\caption{DuMond diagrams for a sequence of four asymmetric Bragg
  reflections from crystals functioning as C-, D$_1$-, D$_2$-, or
  W-element. Anomalous transmission in the F-element is not taken into
  account.  Green and yellow stripes are the regions of Bragg
  reflections in the space of x-ray wavelengths $\lambda$ and angles
  of incidence $\theta_{\indrm{H}}$ or reflection
  $\theta_{\indrm{H}}^{\prime}$ from an H-element (H=C, D$_1$, D$_2$,
  or W).  Blue stripes display the overlapping reflection regions of
  the C- and D-elements. Orange tetragons display the reflection
  region common for all elements.  D-element is set into
  backscattering ($\theta_{\indrm{D}}\rightarrow\pi/2$) with the
  center of the reflection spectral range at
  $\lambda_{\indrm{R}}=hc/E_{\indrm{R}}$ - Eq.~\eqref{eq003}. }
\label{fig004}
\end{figure*}

The presented above qualitative picture is supported by calculation of
the spectral distributions of x~rays after each successive reflection,
based on the dynamical theory of x-ray diffraction in crystals. The
distributions are shown in Figs.~\ref{fig003}(r) and (b). Crystal
parameters used in the calculations are given in Table~\ref{tab1}.
The divergence of the incident beam was assumed to be $20~\mu$rad.
The spectral distribution upon the 1st Bragg reflection from the \cw
-crystal is very broad in agreement with large spectral width of the
$(220)$ Bragg reflection. After the 2nd reflection, the bandwidth is
reduced to the intrinsic width $\dei{H} =27$~meV of the $(800)$ Bragg
reflection from crystal D$_1$. A dramatic change in the spectral
distribution occurs in anomalous transmission through the \cw
-crystal, in the 3rd interaction. In the 4th reflection, from crystal
D$_2$, the spectral distribution practically does not change, as its
Bragg reflection bandwidth is already much broader than the incident
spectral width. The main function of D$_2$ is to increase the angular
dispersion rate by a factor of two, i.e., the opening of the angular
dispersion fan - Eq.~\eqref{eq004}. In the 5th reflection, the \cw
-crystal selects x~rays in a small angular range, and therefore
reduces further the bandwidth to $\Delta
E_{\indrm{\mono}}=0.4$~meV. The peak throughput is $40~\%$.

The evolution of the spectral distributions shows that the most
dramatic change happens in the 3rd step, when in anomalous
transmission the \cw -crystal cuts abruptly the angular dispersion fan
from one side, and reduces it from another. The thicker the crystal,
the steeper the tail is.  It is very close or even steeper than the
slope of the Gaussian function shown by black dashed line in
Figs.~\ref{fig003}(r) and (b).  The effect of anomalous dispersion
resulting in the extremely steep tail on one side is so large, that
the wavelength selector in the 5th reflection improves the spectral
distribution only on the opposite side. Thus, anomalous transmission
is essential for the formation of both small bandwidth and the steep
tails.

The tail can be as steep as that of the Gaussian function, however,
only on one side. For the blue-winged crystal configuration shown in
Fig.~\ref{fig002}(b) the steep tail is on the low-energy side, while in
the other red-winged configuration in Fig.~\ref{fig002}(r), it is on the
high-energy side. Apparently, one can think of an \adat\ optics in
\cdaadw\ or simply \cdaa\ configuration, with two \filter -elements,
which would produce the steep tails on both sides. These options will
be studied elsewhere.

DuMond diagram analysis \cite{DuMond37} provides a valuable graphical
presentation and insight into the complex machinery of the
multi-reflection optics.  The relative spectral bandwidth $\Delta
E_{\indrm{\mono }}/E$, the angular acceptance $\Delta
\theta_{\indrm{\mono }}$ of the monochromator, and the angular
divergence $\Delta \theta_{\indrm{\mono }}^{\prime}$ of x~rays
emanating from the monochromator can be expressed to a good accuracy
in simple terms by
\begin{align}
\frac{\Delta E_{\indrm{\mono }}}{E} & \,=\,\frac{\Delta \lambda_{\indrm{\mono }}}{\lambda}\, =\, \frac{\Delta\theta_{\indrm{C}}^{\prime} +\Delta\theta_{\indrm{W}}}{4\tan\eta_{\indrm{D}}}, \label{eq002a}\\
\Delta \theta_{\indrm{\mono }} &\,=\, \dai{\mathrm C}/ \sqrt{|b_{\indrm{C}}|}, \label{eq002b}\\  
\Delta \theta_{\indrm{\mono }}^{\prime}\,& =\, \dai{\mathrm W} \sqrt{|b_{\indrm{W}}|}.  
\label{eq002c}
\end{align}
They are derived using DuMond diagrams in Fig.~\ref{fig004}, in the
way similar to how they were derived for the CDW-monochromator in
\cite{Shvydko-SB}.  The assignment of reflection regions is given in
the Figure caption.  We note that the inclination of the reflection
region (blue stripe in panel D$_2^{\prime}$-W) representing
wavelength-angular distribution of x rays reflected from the
D$_2$-element is two times greater than the inclination of the
reflection region, which represents the distribution of x rays upon
reflection from the D$_1$-element (blue stripe in panel
D$_1^{\prime}$-D$_2$).  As a result, the bandwidth $\Delta
E_{\indrm{\mono }}$ (Eq.~\eqref{eq002a}) is a factor of two smaller
than the bandwidth of the CDW monochromator with the same crystal
parameters. Equation~(\ref{eq002a}) demonstrates an important
distinguishing feature of the angular dispersive monochromators: the
spectral bandwidth $\Delta E_{\indrm{\mono }}$ is independent of
the intrinsic spectral width of the Bragg backreflection of the D-crystal.
It depends on the strength of the effect of angular dispersion,
expressed by $\tan \eta_{\indrm{D}}$, and it depends on the
geometrical parameters, such as the angular spread
$\Delta\theta_{\indrm{C}}^{\prime } \,=\, \dai{\mathrm
  C}\sqrt{|b_{\indrm{C}}|}$ of the photons emanating from the
collimator crystal C and incident on the dispersing element, and, on
the angular acceptance $\Delta\theta_{\indrm{W}}\,=\, \dai{\mathrm W}
/\sqrt{|b_{\indrm{W}}|}$ of the wavelength selector W.
Another distinguishing feature, the angular acceptance of the \cddw\
optics is determined solely by the angular acceptance of the C-element
- Eq.~\eqref{eq002b}, and it can be made large, more than
$100~\mu$rad, by choosing low indexed Bragg reflections.


\section{Demonstration of  the \adat\ Optics}

Two \cddw\ monochromators, one in blue- and the other in red-winged
configuration, have been designed, built, and commissioned at the
Advanced Photon Source (APS), 30-ID beamline, to study key questions,
whether the \adat\ optics is capable in practice to mold
the spectral distribution of x~rays to profiles with steeply declining
tails, small bandwidths, and can be applied to x-ray beams with large
angular divergence. Parameters of the crystals used in the
monochromators are given in Table~\ref{tab1}. Design monochromator
parameters are given in Table~\ref{tab2}. Technical details on the
experimental set-up, monochromator's mechanical design, crystal
fabrication and characterization, crystal alignment procedure, crystal
temperature control, and other experimental details will be provided
in additional publications \cite{SSG11,SSSK11}.

\begin{table}[t!]
\centering
\begin{tabular}{|l|rrrrrr|}
  \hline 
  & $\Delta E_{\indrm{\mono }}$ & $\Delta E_{\indrm{\mono  \otimes \ana}}$  & $\Delta E_{\indrm{1/10000}}$ & $C_{\indrm{\mono}}$ & $\Delta \theta_{\indrm{\mono }}$ & $\varepsilon_{\indrm{\mono }}$  \\[0pt]    
   \cline{2-7}
  & [meV]  & [meV] & [meV] &   & [$\mu$rad]  &  \%   \\[0pt]    
  \hline  
  Theory     & 0.4  & 0.56 &  1.0 &  1000  &  105  & 22   \\[-0.0pt]
\hline
  Experiment & 0.46 & 0.65 &  3.2  & 500  &  107  & 16   \\[-0.0pt]
  \hline  
\end{tabular}
\caption{Design and measured parameters of the \cddw\ monochromator: $\Delta E_{\indrm{\mono }}$ - 
  full width at half maximum (FWHM) of the spectral resolution function 
  of a single monochromator; $\Delta E_{\indrm{\mono  \otimes \ana}}$ - 
  FWHM for a combined spectral resolution functions of the blue-winged 
  monochromator and the red-winged analyzer;   $\Delta E_{\indrm{1/10000}}$ -  
  spectral half width at the $10^{-4}$ level fraction
  of the maximum  on the side of the spectral resolution function with steeper tail, 
  $C_{\indrm{\mono }}$  - spectral contrast,
  $\Delta \theta_{\indrm{\mono }}$ angular acceptance, and $\varepsilon_{\indrm{\mono }}$ average
  spectral efficiency. The latter is defined as 
  $\varepsilon_{\indrm{\mono }}=I_{\indrm{\mono }}/I_{\indrm{0}}\times \Delta
  E_{\indrm{0}}/\Delta E_{\indrm{\mono }}$, with  $I_{\indrm{0}}$  as incident and $I_{\indrm{\mono }}$ 
  transmitted through the monochromator photon flux, 
  and $\Delta E_{\indrm{0}}$  as FWHM of the incident radiation spectral distribution. }
\label{tab2}
\end{table}

\begin{figure}[h!]
\setlength{\unitlength}{\textwidth}
\begin{picture}(1,0.96)(0,0)
\put(0.0,-0.02){\includegraphics[width=0.5\textwidth]{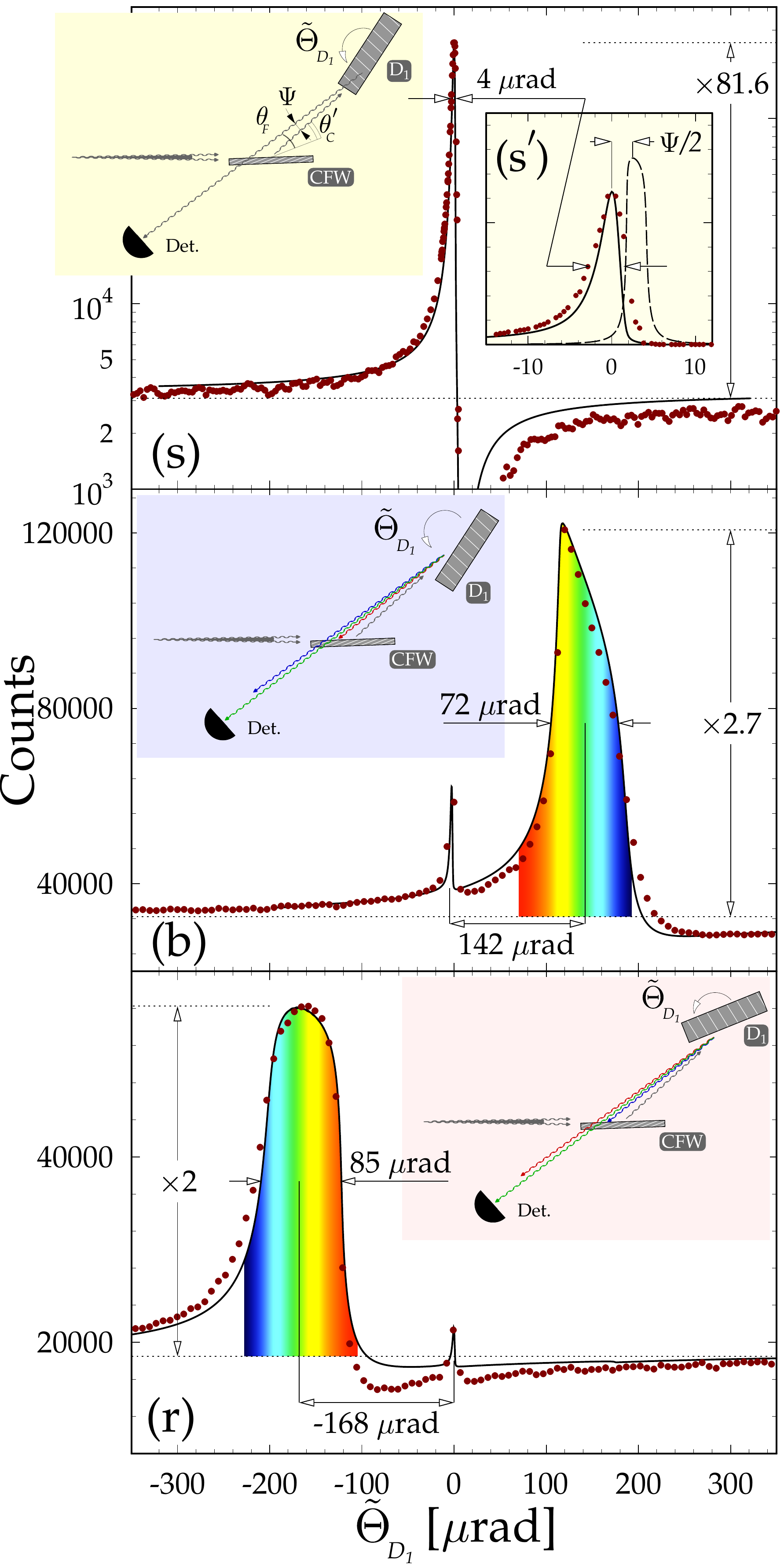}}
\end{picture}
\caption{Angular distributions of x-ray intensity upon a sequence of
  interactions C$\rightarrow$D$_1\rightarrow$F, schematically shown in
  the insets, representing: (s) effect of anomalous transmission, and
  (b), (r) combined effect of angular dispersion and anomalous
  transmission of x~rays in Bragg diffraction from asymmetrically cut
  crystals. Sharp line ($\simeq 4~\mu$rad) in (s) is due to
  backreflection and subsequent anomalous transmission taking place
  for all photon energies at the same incidence angle to crystal D$_1$
  (same rotation angle $\tilde\Theta_{\indrm{D_1}}=0$). In contrast,
  in (b) and (r) backreflection takes place at different angles for
  different photon energies, indicated by color.  Solid and dashed
  lines are dynamical theory calculations, and solid circles are
  experimental data.  (s$^{\prime}$) shows same dependences as (s) but
  on the linear scale.  }
\label{fig005}
\end{figure}

Perfect implementation of the angular dispersion and anomalous
transmission effects is critical for achieving the anticipated
performance of the \adat\ optics. Measurements presented in
Fig.~\ref{fig005}, and discussed in the following, demonstrate that both
effects perform close to theoretical expectations. The studies are
performed in the CD\filter\ configuration as shown schematically by
the scattering diagrams in the insets of Figure~\ref{fig005}.  The beam
is collimated to $\simeq 1~\mu$rad divergence after the first
reflection from the \cw -crystal, monochromatized to $\simeq 27~$meV
bandwidth upon backreflection from the D$_1$-crystal, then transmitted
through the \cw- crystal, and recorded using a photon counting
detector (Det.)  as a function of $\rotangleone$, the angular
coordinate of the D$_1$-crystal. The variation of the rotation angle
$\delta\rotangleone $ directly relates to the variation of the
incidence angle to \cw\ by $\delta \theta_{\indrm{\filter
  }}=2\delta\rotangleone $.

\subsection{Anomalous Transmission}

Results presented in Figs.~\ref{fig005}(s) and (s$^{\prime}$) are
obtained when the lateral face of the D$_1$ crystal is illuminated.
For simplicity, we refer to this configuration as {\em symmetric},
since the reflecting atomic planes compose small asymmetry angle with
the lateral face.  The recorded intensity (Counts) is shown by solid
circles in Figs.~\ref{fig005}(s) on the logarithmic scale.  A sharp
asymmetric transmission peak is observed with the position of the
maximum chosen at $\rotangleone=0$.  The peak value exceeds by a
factor of $\gtrsim 80$ the normal level of transmission, indicated by
the lower horizontal dotted line.  This is an enhancement of the
anomalous transmission effect an order of magnitude greater than the
largest previously reported in literature \cite{BHS75}. The
experimental curve is overall in a good agreement with the theoretical
dependence shown by the solid black line, though the angular width of
$4~\mu$rad is somewhat broader than $2.5~\mu$rad expected in
theory. It has a typical for the angular dependence of anomalous
transmission dispersion form with a very steep edge changing over into
a broad minimum on the positive $\rotangleone$ side.

Figure~\ref{fig005}(s') shows the same two dependences on the linear
scale.  In addition, the dashed line represents a calculated angular
dependence of the accompanying Bragg reflection from the \cw\ crystal
which was not measured in the experiment.  It shows that, the
reflection peak is shifted by only $\rotangleone =\Psi/2$
($\Psi=5~\mu$rad) with respect to the transmission peak.  Thus, in
agreement with expectations, the \cw\ crystal transmits x~rays only at
an angle of incidence $\theta_{\indrm{\filter }}
=\theta_{\indrm{C}}^{\prime}-\Psi$, which differs from the glancing
angle of reflection $\theta_{\indrm{C}}^{\prime}$.  In other words,
the \cw\ crystal transmits x~rays which are only in almost exact
backreflection from D$_1$ crystal.  The experimental facts demonstrate
that the effect of anomalous transmission is perfectly working.

\subsection{Angular Dispersion and Anomalous Transmission}

Figures~\ref{fig005}(r) and (b) show results of similar measurements,
however, with the x-ray beam reflected from the asymmetrically cut
face of the D$_1$ crystal, in the red- and blue-winged configurations,
respectively. Unlike the previous case, the anomalous transmission
takes place in a much broader $\simeq 72-85~\mu$rad angular range of
the rotation angle $\rotangleone $, and the maximum is observed at $
\rotangleone \simeq 142~\mu$rad (b) and at $\rotangleone \simeq
-168~\mu$rad (r), respectively.  (The weak sharp peaks are of the same
nature and at the same angular position $\rotangleone =0$ as the peak
in Fig.~\ref{fig005}(s), appearing because a small part of the incident
beam still illuminates the lateral crystal face of the D$_1$ crystal.)
Nothing has changed, compared to the previous case, with the condition
for anomalous transmission through the \cw\ crystal: it takes place
only at an angle of incidence $\theta_{\indrm{\filter }}
=\theta_{\indrm{C}}^{\prime}-\Psi$.  What has dramatically changed is
the backreflection condition, by transition to diffraction from the
asymmetrically cut crystal face.

Now, in asymmetric diffraction - Figs.~\ref{fig005}(b) and (r),
backreflection from D$_1$ takes place, in agreement with
Eq.~\eqref{eq003}, at different angles for different photon energies
as indicated by the color, i.e., in a broad angular range, which peak
is also shifted by $\Theta_{\ind{R}}$.
Using Eq.~\eqref{eq003} and the crystal parameters from
Table~\ref{tab1}, we obtain $\Theta_{\ind{R}}\, \simeq
168~\mu$rad. The angular width of the exact Bragg backscattering from
crystal $D_1$ with intrinsic spectral width $\dei{\mathrm
  D_1}=27$~meV, can be estimated using Eq.~\eqref{eq003} as
$\Delta\Theta \,=\, (\dei{\mathrm
  D_1}/E_{\ind{R}})\tan\eta_{\indrm{D_1}}\simeq 85~\mu$rad.  The
measured values shown in Fig.~\ref{fig005}(b) agree well, while values
shown in Fig.~\ref{fig005}(r) perfectly, with these estimations.  The
interplay between anomalous transmission and angular dispersion, which
is not taken into account in this estimation, is different in the two
cases, resulting in different shapes, positions, widths, and
amplitudes of the peaks. Rigorous multi-crystal dynamical theory x-ray
diffraction calculations shown by solid lines are in good agreement in
both cases, with regard to the width, position, amplitude of anomalous
transmission, and shape of the exact backscattering peak.
These observations demonstrate an overall good functioning of both
effects of angular dispersion and anomalous transmission.

\subsection{\cddw\ Angular Acceptance}

\begin{figure}[t!]
\setlength{\unitlength}{\textwidth}
\begin{picture}(1,0.51)(0,0)
\put(0.0,0.0){\includegraphics[width=0.5\textwidth]{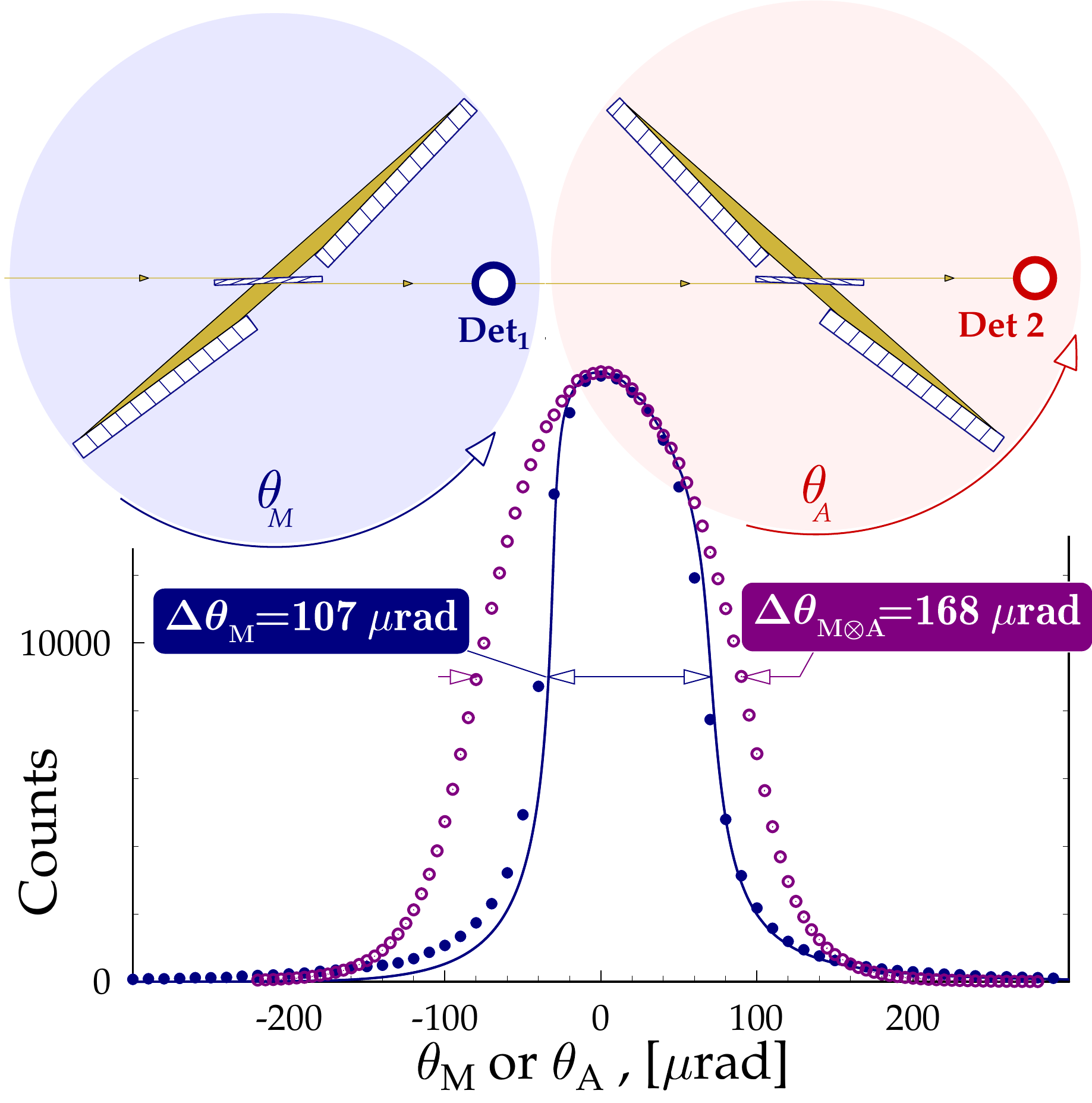}}
\end{picture}
\caption{Experimental scheme and angular dependence of transmission
  through \cddw\ monochromators. Solid blue circles show results for the
  blue-winged monochromator with incident beam collimated to $\simeq
  15~\mu$rad, as measured by Det$_1$ x-ray detector. Solid line -
  theory. Open purple circles: show results for the red winged analyzer
  with the incident beam emanating from the monochromator, set to the
  transmission maximum, as measured by Det$_2$ x-ray detector.  }
\label{fig006}
\end{figure}

\begin{figure}[t!]
\setlength{\unitlength}{\textwidth}
\begin{picture}(1,0.51)(0,0)
\put(0.0,0.0){\includegraphics[width=0.5\textwidth]{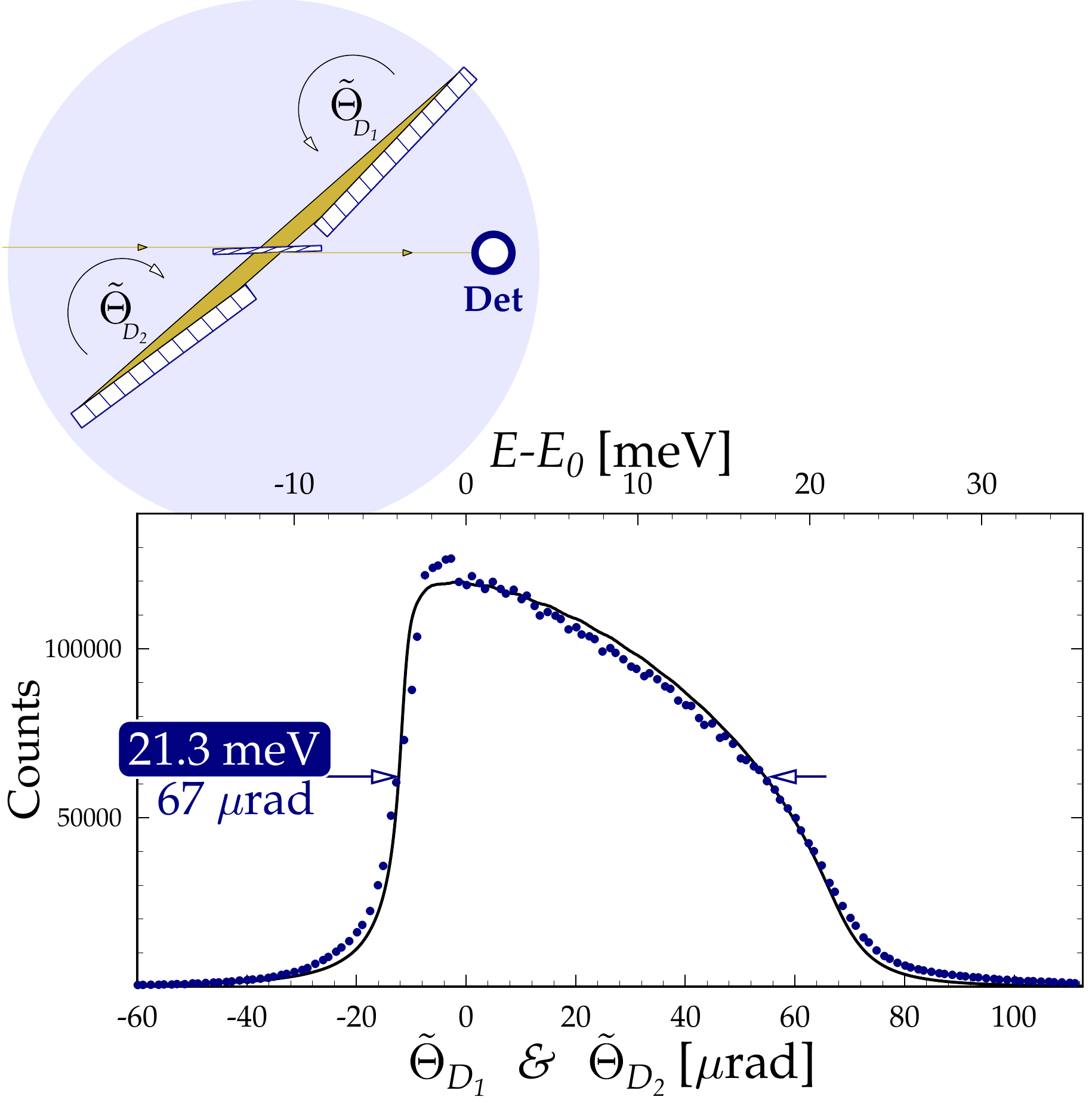}}
\end{picture}
\caption{Experimental scheme and a spectral tuning curve of the \cddw\
  monochromator measured by rotation of the D-crystals. The blue
  circles show experimental data, while the solid line - calculations
  using multi-crystal dynamical theory of x-ray diffraction.}
\label{fig007}
\end{figure}

\begin{figure}[t!]
\setlength{\unitlength}{\textwidth}
\begin{picture}(1,0.65)(0,0)
\put(0.0,0.0){\includegraphics[width=0.5\textwidth]{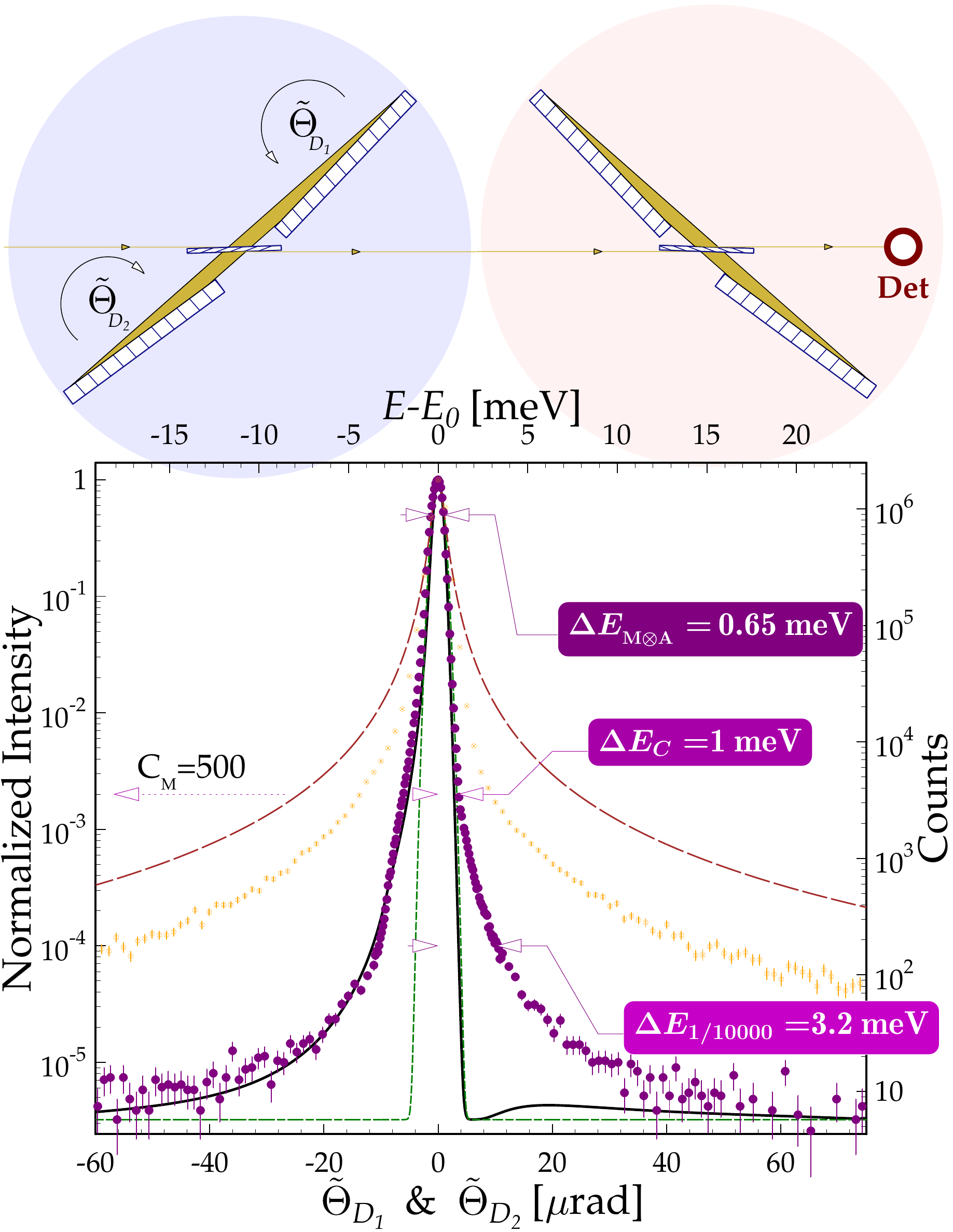}}
\end{picture}
\caption{Experimental scheme and combined spectral resolution function
  of the blue-winged \cddw\ monochromator measured against the
  red-winged analyzer. The purple circles show experimental spectral
  resolution function, the black solid line - spectral functions
  calculated using multi-crystal dynamical theory of x-ray
  diffraction. Other functions with the same FWHM are shown for
  comparison: green dashed line - Gaussian; dark-red dashed line -
  Lorentzian, orange open circles - experimental dependence for a
  four-crystal monochromator \cite{Chumakov11}.}
\label{fig008}
\end{figure}

The two \cddw\ monochromators were aligned, one in blue- , another in
red-winged configuration - Figs.~\ref{fig006}-\ref{fig008}
\footnote{The red-winged \cddw\ optics shown in the schemes of
  Fig.~\ref{fig002}(r) and Figs.~\ref{fig006}, \ref{fig008} are
  equivalent. In the experiments we realized the scheme shown in
  Figs.~\ref{fig006}, \ref{fig008} as it offered better possibilities
  for strain-free crystal mount \cite{SSSK11}.}. Their performance was
evaluated in terms of angular acceptance, widths and contrast of the
spectral function, and the average spectral efficiency. Measured
values as well as design parameters are given in Table~\ref{tab2}.
The values are obtained from the appropriate angular and spectral
dependences shown in Figs.~\ref{fig006}, \ref{fig007}, and
~\ref{fig008}, respectively, along with the schemes of the
experimental arrangements.

The angular acceptance has been measured by rotation of a \cddw\
monochromator, and by counting photons arriving at its exit after the
five reflections. Blue circles in Fig.~\ref{fig006} represent the
measured dependence, while the solid line represent the results of the
multi-crystal dynamical theory calculations for the incident beam with
a $\simeq 15~\mu$rad divergence, and a $\Delta E_{\indrm{0}}=0.6$~eV
bandwidth, as in the experiment. There is a very good correspondence
between the measured and calculated dependences, with an angular
acceptance of $\Delta\theta_{\indrm{\mono }}=107~\mu$rad.  This is an
unusually large number. Typically, the angular acceptance of
high-resolution x-ray monochromators is in a $10$ to $20~\mu$rad
range, often requiring collimating optics, such that the x~rays from
the source to be fully accepted by the monochromator
\cite{ChM96,Toellner00,CRL01,Baron01,Yabashi01,Toellner01,TAS06,TAS11}.
The only exception to this rule are single bounce monochromators
\cite{VSK96}, which, however, have a disadvantage of long Lorentzian
tails in the spectral resolution function.

In the second measurement presented by the purple open circles, the
beam from the first monochromator (set to the transmission maximum) is
guided through the second monochromator, which is rotated about its
axis. The width of the angular dependence in this measurement is
$\simeq 168~\mu$rad, about $50\%$ broader than the angular width
measured with the direct beam. This agrees with the expectation that the
angular divergence of x~rays from the \cddw\ monochromator $\Delta
\theta_{\indrm{\mono }}^{\prime}$ has to be as large as the angular
acceptance $\Delta \theta_{\indrm{\mono }}$ -
Eqs.~\eqref{eq002b}-\eqref{eq002c}, and therefore the broader angular
curve is a result of the convolution of a $\simeq 100~\mu$rad
divergent beam with $\simeq 107~\mu$rad angular acceptance of the
monochromator.

\subsection{\cddw\ Spectral Resolution Function}


The peak photon energy of the monochromator spectral resolution
function can be tuned by changing simultaneously the angles of
incidence $\theta_{\indrm{D_1}}$ and $\theta_{\indrm{D_2}}$
(Fig.~\ref{fig002}), as formally expressed by Eq.~\eqref{eq003}. This
can be accomplished through simultaneous rotation of the D-crystals
with rotation angles $\rotangleone$ and $\rotangletwo$, as shown in
the scheme of Fig.~\ref{fig007}.  The relationship between the angular
and energy variations in our case $\delta
E/\delta\Theta_{\indrm{D}}=0.319$~meV/$\mu$rad, which is obtained from
Eq.~\eqref{eq005}, and crystal parameters in Table~\ref{tab1}.

Solid dots in Fig.~\ref{fig007} represent thus measured spectral tuning
curve of the \cddw\ monochromator.  The solid line shows the results
of the theoretical calculations, which is in a good agreement with the
experimental data.  The width of the tuning curve is related to the
intrinsic width $\dei{\mathrm D}=27$~meV of the D-crystal reflection
curve. The measured tuning curve width ($\simeq 21$~meV) is, however,
smaller, because it represents the width of the product of two Bragg
reflection curves resulting from the sequence of two
backreflections. The tuning curve shows that the available tuning
range is relatively small, if the energy is changed by rotation of
D-crystals. The tuning range can be increased by varying the
temperature of D-crystals \cite{SKR06}, which would result in the
change of the lattice parameter $\dhkl$ and backscattering energy
$E_{\ind{R}}$ - Eq.~\eqref{eq003}.

The \cddw\ monochromator spectral resolution function was measured by
changing the monochromator energy (by rotation of D-crystals), and
using another monochromator as an analyzer, as shown in the scheme of
Fig.~\ref{fig008}.  Convolution of the spectral resolution functions of
the monochromator and the analyzer, a combined spectral resolution
function, is measured in this case. The expected \cddw\ spectral
function is asymmetric with the very steep tail on one side -
Figs.~\ref{fig003}(r) and (b).  For the combined spectral resolution
function to preserve the steep tail, the monochromator and analyzer
have to be chosen, such that one is in the blue- and the other in the
red-winged configuration. The steep tail of the combined spectral
resolution function will be reproduced on the side where the spectral
function of the device which is not tuned has the steep tail. In our
experiment, the blue-winged monochromator is tuned, and the red-winged
analyzer is at a fixed energy.

The purple solid circles in Fig.~\ref{fig008} show the results of the
measurements of the combined spectral resolution function on the
logarithmic scale. An x-ray detector is installed downstream the
analyzer.  The full width at half maximum is $\Delta E_{\indrm{\mono
    \otimes \ana}}=0.65$~meV, which is close to the design value of
0.56~meV. It is important to note, that this very high resolution is
achieved with an x-ray beam incident upon the analyzer, that has a
large angular divergence of $\Delta \theta_{\indrm{\mono
  }}^{\prime}\simeq 100~\mu$rad.  Most importantly, the function has a
steeply declining tails especially on the high-energy side with
spectral contrast $C_{\indrm{\mono }} \simeq 500$, and a half width at
the $10^{-4}$ level fractions of the maximum $\Delta
E_{\indrm{1/10000}}=3.2$~meV. Theory predicts even steeper tail, shown
by the black solid line in Fig.~\ref{fig008} with $C_{\indrm{\mono }}
\simeq 10^3$, and $\Delta E_{\indrm{1/10000}}=1.0$~meV. The
discrepancy is attributed to yet not fully perfect sub-surface layer
of silicon crystals used in the experiment. Improvements in crystal
fabrication are in progress \cite{SSSK11}. Nonetheless, we are
measuring a spectral function with remarkably steep tails. First, the
experimental curve follows the tails of the Gaussian function over
almost three orders of magnitude. Secondly, the measured curve is more
than an order of magnitude steeper than the tails of the best spectral
resolution functions measured with the state-of-the-art multi-crystal
x-ray optics. The orange circles show an example of such resolution
function for a four-crystal monochromator designed for nuclear
resonant scattering experiments with 14.4 keV photons
\cite{Chumakov11}. It has been measured using a very well collimated
($\simeq 3~\mu$rad) incident x-ray beam, and a $\simeq 50$-neV-broad
nuclear resonance as an analyzer. Thirdly, the tails of the combined
\cddw\ spectral resolution function is two orders of magnitude steeper
than the tails of the Lorentzian distribution with the same FWHM. It
should be noted that the long Lorentzian tails (spectral contrast
$C_{\indrm{\mono }}=11$), are typical for spectral resolution functions
of all existing inelastic x-ray scattering (IXS) spectrometers
\cite{MBKRSV96,Baron1,Sinn01,SSD11}.
For the Lorentzian tails to reach the level of the measured \cddw\
resolution function, the width of the Lorentzian distribution would
have to be reduced to $65~\mu$eV. The latter has to be compared with a
$1.5$~meV width (a more than 20 times larger value) of the spectral
resolution functions presently available with the state-of-the-art IXS
spectrometers.

The last but not least, the measured average spectral efficiency
$\varepsilon_{\indrm{\mono }}=16~\%$, is close to $22~\%$ expected in
theory (see Table~\ref{tab2} caption for the definition of
$\varepsilon_{\indrm{\mono }}$).


In conclusion, the angular dispersive and anomalously transmissive x-ray
(\adat ) optics offer a possibility to shape x-ray spectra to
distributions with steeply declining tails (large contrast), and
small bandwidth.  The \adat\ optics is applicable to x~rays with a
large spread of angles of incidence.  Introduced here, a
three-crystal, five-reflection \adat\ x-ray optics, the \cddw\
monochromators feature a combination of superlative properties:
exceptionally steep tails of the spectral profile, extremely narrow
bandpass, extraordinary large angular acceptance, high efficiency, and
in-line configuration. The monochromators and analyzers based on this
principles have a potential to become key optical components in the
next generation ultra-high resolution inelastic x-ray scattering
spectrometers, and other applications in x-ray science.



\section{Acknowledgments}

We are grateful to Linda Young for supporting this project in its
decisive phase at the APS.  We are indebted to J.-H.~Kim (APS),
X.~Huang (APS), and J.~Sutter (DLS), for help with the experiments at
the APS 30-ID beamline.  T.~Roberts (APS) is acknowledged for long
standing technical support, K.~Goetze (APS) for developing
monochromator controls, M.~Wiczorek for help with crystal fabrication,
L.~Assoufid, Q.~Qian, and J.~Mai with metrology, K.~Mundboth (DLS) for help
with developing temperature control.  A.I.~Chumakov (ESRF) is
acknowledged for providing unpublished data.  S.P.~Collins (DLS),
C.~Burns (WMU), Y.~Cai (BNL), S.-H.~Chen (MIT), J.P.~Hill (BNL),
B.C.~Larson (ORNL), G.~Monaco (ESRF), D.~Reznik (Uni. of Colorado at
Boulder), G.~Ruocco (Sapienza University of Rome), Q.~Shen (BNL), and
T.~Rayment (DLS) are acknowledged for stimulating interest and
valuable suggestions. Work was supported by the U.S. Department of
Energy, Office of Science, Office of Basic Energy Sciences, under
Contract No. DE-AC02-06CH11357.
%

\end{document}